\documentclass[twocolumn,pra,aps,superscriptaddress,showpacs,amsmath,amssymb,floatfix,longbibliography]{revtex4-1}
\usepackage{graphicx}
\usepackage{amsmath}
\usepackage{siunitx}
\usepackage{bbm}
\usepackage{float}
\usepackage[colorlinks,
linkcolor=red,
anchorcolor=blue,
urlcolor=blue,
citecolor=green]{hyperref}

\begin{document}

%\preprint{}

\title{High-precision gravimeter based on a nano-mechanical resonator hybrid with an electron spin} %Title of paper



\author{Xing-yan Chen}
\affiliation{School of Physics, Peking University, Beijing 100871, China}
\affiliation{Center for Quantum Information, Institute for Interdisciplinary Information Sciences, Tsinghua University, Beijing 100084, China}

\author{Zhang-qi Yin}\email{yinzhangqi@tsinghua.edu.cn}
\affiliation{Center for Quantum Information, Institute for Interdisciplinary Information Sciences, Tsinghua University, Beijing 100084, China}


\date{\today}

\begin{abstract}
We show that the gravitational acceleration can be measured with the matter-wave Ramsey interferometry, by using a nitrogen-vacancy center coupled to a nano-mechanical resonator. We propose two experimental methods to realize the Hamiltonian, by using either a cantilever resonator or a trapped nanoparticle.  The scheme is robust against the thermal noise, and could be realized at the temperature much higher than the quantum regime one.
The effects of decoherence on the interferometry fringe visibility is caculated,  including the mechanical motional decay and dephasing of the  nitrogen-vacancy center. In addition, we  demonstrate that under the various sources of random and systematic noises, our gravimeter can be made on-chip and achieving a high measurement precision. Under experimental feasible parameters, 
the proposed gravimeter could achieve  $10^{-10}$ relative precision.
\end{abstract}


\maketitle %\maketitle must follow title, authors, abstract and \pacs


\section{Introduction}

High quality nano(micro)-mechanical resonator is one of the best testbed for fundamental physics \cite{Aspelmeyer2014a}, such as the macroscopic quantum superpositions \cite{Chang2010,Romero2010}, the gravity induce wavefunction collapse \cite{Bassi2013},  the boundary between quantum and classical regimes \cite{Arndt2014,Yin2017}, and etc. It is found that the large quantum superpositions of the nano-mechanical resonator could be realized with the help of cavity modes \cite{Romero2011}, superconducting circuits \cite{Li2016,Khosla2018}, nitrogen-vacancy centers \cite{Rabl2009,Yin2013,Scala2013b,Ma2017}, and etc. 
The quantum-classical boundaries can be tested in these systems through matter-wave 
interferometry \cite{Li2016,Bose2017,Marletto2017b}. On the other hand, the nano(micro)-mechanical resonator is also
widely used in precision measurement of masses \cite{Zhao2014}, torsion \cite{Hoang2016a}, forces and accelerations \cite{Ranjit2016,Monteiro2017}, because of its high mechanical Q. 

As we know, the interferometry firstly appeared in optics, which was used for prescise measurement.
Later, the matter-wave interferometry was realized with electrons \cite{Merli1976} and neutrons \cite{Rauch1974}, then with larger particles such as atoms \cite{Carnal1991,Monroe1996} and molecules \cite{Hornberger2012}. Atom interferometry has evolved from the demonstration of quantum superpositions into instruments at the cutting edge of precise measurement, including measurements of platform rotation, the Molar-Planck constant, the fine structure constant \cite{Godun2001,Cronin2009} and the gravitational acceleration \cite{Peters2001,Bodart2010,Zhou2011,Zhou2012}. One of the motivations to replace the light with atoms for interferometry is that the shorter atomic de Broglie wavelength could make the measured phase shift much more accurate. Therefore it is 
natural to anticipate that the interferometers with macroscopic object, such as the nano-mechanical resonators, could greatly increase the  measurement precision of the phase shift. 

In this paper, we propose a scheme to realize the high-precision gravimeter with a nano-mechanical 
resonator hybrid with a nitrogen-vacancy center by using matter-wave interferometry. We give a physically intuitive derivation of the interferometer phase shift. 
With the state-of-the-art technologies, we estimate the phase shift to be 3 orders of magnitude larger than those using atom interferometry \cite{Peters2001,Cronin2009,Hornberger2012}. 
We briefly analyze random and systematic noise and find that the relative measurement precision $10^{-10}$ for gravitational acceleration is achievable.
 Besides, our scheme is solid based and on-chip. Unlike
the gravimeter based on atomic interferometry, here neither the complex lasers nor the
big vibration isolation system is required. Therefore, our scheme is suitable for  portable gravimeters with a high precision.

\section{The scheme}
\begin{figure}[htbp]
	\centering
	\includegraphics[width=0.47\linewidth]{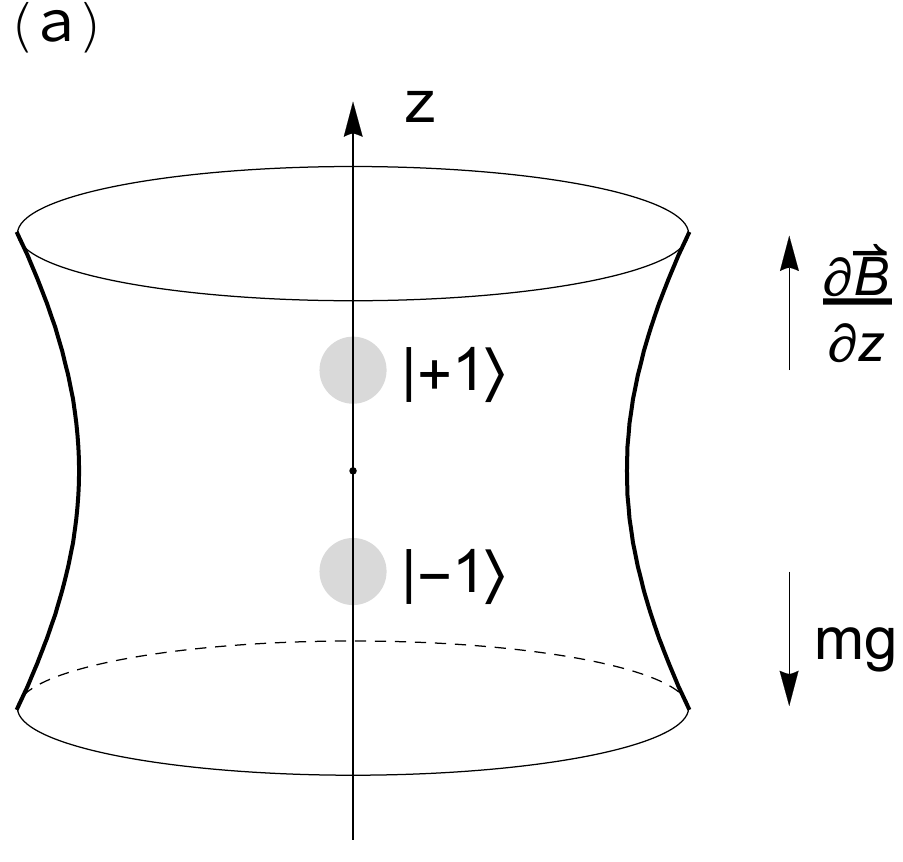}
	\includegraphics[width=0.48\linewidth]{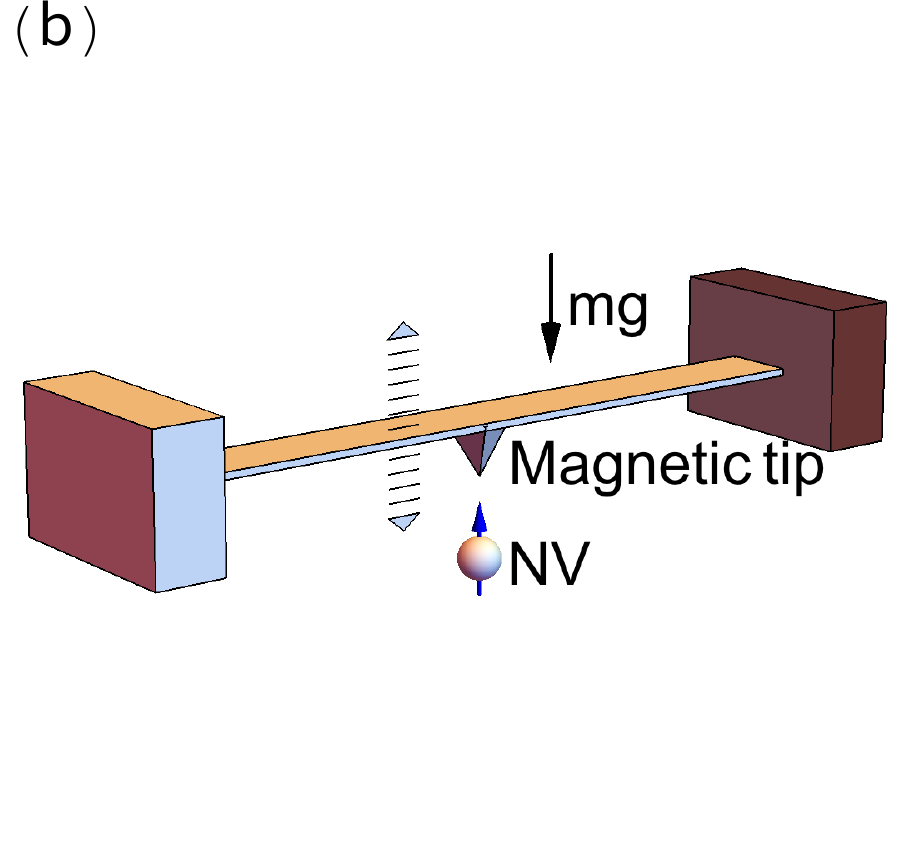}
	\caption{(a) An optical trap holds a nano-diamond  with a build-in NV center with both the weakest confinement and the electron spin quantization along the $z$ axis. A magnetic gradient along the $z$ axis produces spin-dependent shifts to the center of the harmonic well. The $z$ axis is oriented along the vertical gravitational acceleration by tuning the control system. 
%Thus place the centers of $|\pm1\rangle$ states wavefunctions $z_{\pm}$ in the different gravitational potentials. 
The CoM of the nano-diamond oscillates around the two balanced points $z_{\pm}$, accumulating a relative gravitational phase difference $\Delta \phi$. At $t_0 = 2\pi/\omega_z $ this phase can be read from spin population.
	(b) A scheme to strongly couple an NV center with a cantilever. The NV center is embedded in bulk diamond lattice. A magnetic tip is attached to the cantilever, which provides the magnetic gradient. This setup can fulfill our requirement of large magnetic field gradient and long dephasing time $T_2$, thus can significantly improve gravitational measurement precision.}
	\label{fig:trap}
\end{figure}

We consider a nano-mechanical resonator hybrid with a nitrogen-vacancy (NV) center through magnetic field gradient induced coupling.
There are two different setups.
The first one, as shown in Fig.\ref{fig:trap}(a), consists of a nanoscale diamond bead containing a  NV center levitated by an optical tweezer in ultrahigh vacuum \cite{Yin2013b,Neukirch2015,Neukirch2015a,Hoang2016}.
  A magnetic tip nearby induces a large magnetic field gradient, and couples the NV center with the
center of mass (CoM) motion of the nano-diamond. 
The other setup uses a cantilever resonator \cite{Rabl2009,Xu2009,Rabl2010,Arcizet2011,Kolkowitz2012,Yin2015}.
As shown in Fig.\ref{fig:trap}(b), a magnetic tip attached to the cantilever is used to couple the mechanical mode to an NV center embedded in bulk diamond bellow the cantilever. In both setups,  the mechanical motion can be described by the same Hamiltonian \eqref{eq:H}, as we will discuss bellow.
 In both cases, the gravity induced dynamical phase is measured through a Ramsey scheme similar to atom interferometry \cite{Kasevich1991a,Peters2001,Ramsey2014,Muller2010}.
We will analyze the phase shift of an oscillator in the gravitational field coupled to a solid spin, then this phase shift is revealed by Ramsey interferometry to measure the gravitational acceleration. 
%Our scheme can also be achieved with a cantilever \cite{Rabl2010,Xu2009}. 

As an example, we analyze the optically levitated nano-diamond scheme in detail. 
 The motion of the nano-diamond in the harmonic potential of the tweezer is coupled to the $S=1$ spin of the NV center by a magnetic field gradient $B_g = \partial B/\partial z$ oriented along the $z$ direction, which can be generated by a magnetic tip. We assume the trapping frequencies satisfy $\omega_x,\omega_y \gg \omega_z$. Therefore the effects of motion on x and y dirctions are neglected.  Then 
the whole system Hamiltonian, including the Earth's gravitational field effects, reads  \cite{Scala2013b}
\begin{equation}\label{eq:H}
	H = \hbar DS_z^2 + \hbar \omega_z c^\dagger c - 2(\lambda S_z - \Delta\lambda)(c + c^\dagger),
\end{equation}
where  $D = \SI{2.88}{GHz}$, $\lambda = g_{NV}\mu_B B_g \sqrt{\hbar/2m\omega_z}$ and $\Delta\lambda = \frac{1}{2}mg\sqrt{\hbar/2m\omega_z}$. The Hamiltonian \eqref{eq:H} represents a harmonic oscillator whose CoM motion depends on the eigenvalue of $S_z$. We denote the eigenstates with eigenvalue $S_z = -1,0,1$ as $|-1\rangle$, $|0\rangle$, and $|+1\rangle$, respectively. For each state we can calculate the evolution of the oscillator. Here we use the Feynman path integral approach \cite{Storey1994}, following the discussions in Ref. \cite{Peters1997} on atomic interferometry.

\begin{figure}[htbp]
	\centering
	\includegraphics[width=\linewidth]{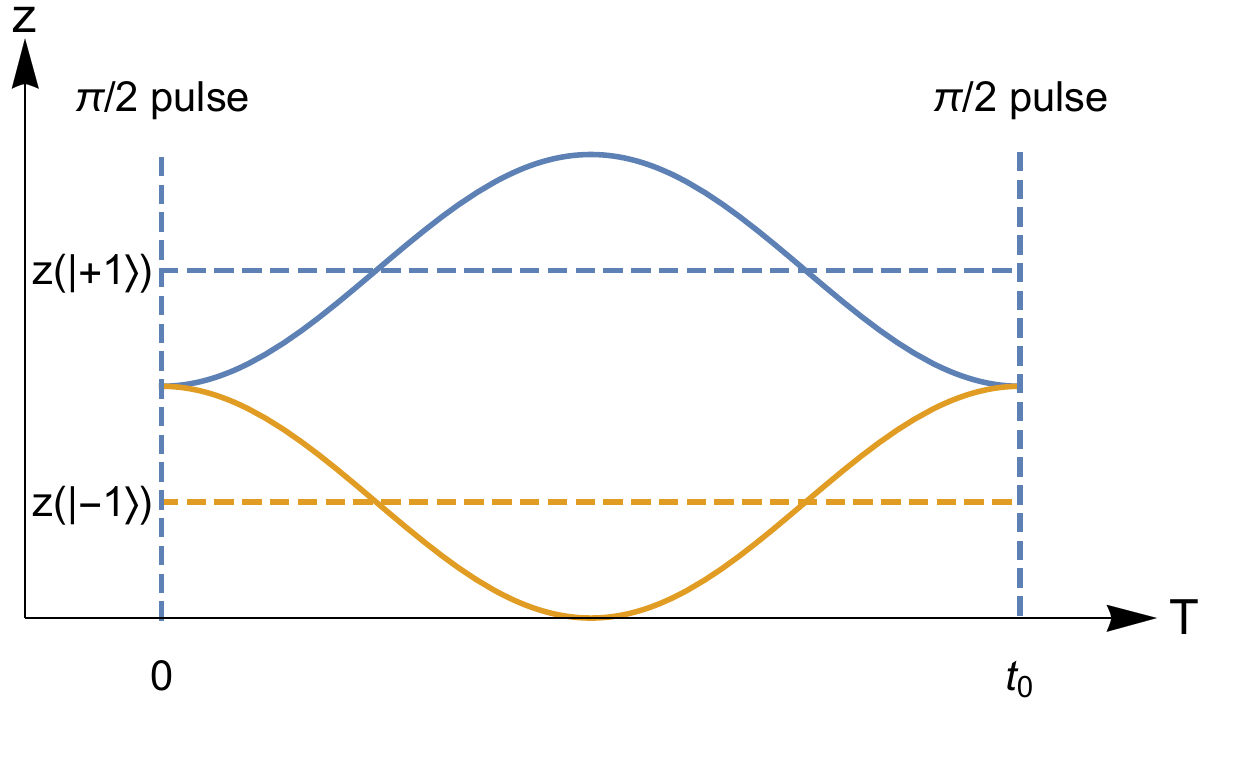}
	\caption{Phase space diagram of the matter wave interferometer based on $\pi/2$ - $\pi/2$ pulse sequence. The nano-object can either be in the internal NV spin state $|+1\rangle$ (blue) of $|-1\rangle$ (orange). The lines represent the classical trajectories originating from one of the space-time points comprising the initial wave packet.}
	\label{fig:path2}
\end{figure}

The phase diagram of Ramsey interferometry based on $\pi/2 - \pi/2$ pulse sequence is shown in Fig.~\ref{fig:path2}. We assume that the NV center is initialized to state $|0\rangle$, and CoM motion of the nano-diamond is cooling down to mK or lower. In the first step, we apply a microwave pulse corresponding to the effective interaction Hamiltonian $H_{\mathrm{mw}} = \hbar \Omega (|+1\rangle\langle0| + |-1\rangle\langle0| + \mathrm{H.c.})$, where $\Omega$ is the Rabi frequency.
In the limit that $\Omega$ is much larger than any other coupling strength in Eq. \eqref{eq:H}, we can neglect any other interactions when applying the pulse. With the pulse duration $t_p = \pi/(2\sqrt{2}\Omega)$, the NV center electron spin state becomes $(|+1\rangle + |-1\rangle)/\sqrt{2}$, which is a superposition of $|+1\rangle$ and $|-1\rangle$ with equal amplitudes. The two states experience a different force due to the spin-dependent coupling term $2\lambda S_z (c+c^\dagger)$, which leads to an additional spin-dependent acceleration
\begin{equation}
g_{\pm} = \pm\frac{g_{NV}\mu_B}{2m}B_g.
\end{equation}  
The equilibrium position of the two states is then determined by $z_\pm = z_0 \pm \Delta z$ as shown in Fig.\ref{fig:path2}, where $z_0 = g/\omega_z^2$ and the displacement
\begin{equation} \label{eq:amplitude}
\Delta z = |g_{\pm}|/\omega_z^2.
\end{equation}
So after the $\pi/2$ pulse the two states will oscillate around their equilibrium points $z_{\pm}$ and the two paths will recombine after an oscillation period $t_0 = 2\pi/\omega_z$. The spin states after the oscillation period is $(|+1\rangle + e^{i\Delta\phi}|-1\rangle)$. The phase shift between the two paths due to propagation can be calculated by their classical actions
\begin{equation}\label{eq:action}
	\Delta \phi = S[z_+(t),t_0,g_+] - S[z_-(t),t_0,g_-],
\end{equation}
with $S[z(t), t_0, g]$ denotes the action over the classical path $z(t)$ of a spring oscillating for one period in the gravitational field. The classical action is given by
\begin{equation}
\begin{split}
& S[z(t),t_0,g] = \int^{t_0}_{0} L(z,\dot{z})dt, \\
& L(z,\dot{z})=\frac{1}{2}(m\dot{z}^2 - \omega_z^2 z^2) + mgz. \\
\end{split}
\end{equation}
Evaluating the integral over the classical path of the oscillator, where the paths of two spin states
\begin{equation}
	z_{\pm}(t) = \pm \Delta z (1-\cos (\omega t)) + z_0,
\end{equation}
 we get the phase shift 
\begin{equation}\label{eq:phase}
	\Delta \phi = \frac{16\lambda\Delta\lambda}{\hbar^2\omega_z}t_0 = \frac{g_{NV}\mu_B}{\pi^2 \hbar}B_g g t_0^3.
\end{equation}
To reveal $\Delta \phi$ we apply another $\pi/2$ pulse  $H_{\mathrm{mw}} = \hbar \Omega (|+1\rangle\langle0| + e^{i\phi}|-1\rangle\langle0|) + \mathrm{H.c.}$ with a relative phase $\phi$. After time $t_p$, the population of the spin state with $S_z = 0$ becomes
\begin{equation}
	P_0(t = t_0 + t_p) = \cos^2\big( \frac{\Delta\phi + \phi}{2} \big),
\end{equation}
which depends on both the phase shift $\Delta \phi$ and relative phase $\phi$ in the second $\pi/2$
pule. The relative phase $\phi$ is scanned to reveal the interference fringes. In this way, the gravity induced the phase offset  $\Delta \phi$ can be precisely measured, leading to precision measurement of the gravitational acceleration $g$.
The above discussion is still valid when the nano-mechanical resonator is in thermal state \cite{Scala2013b}. Therefore we conclude that the interference is immune to thermal motion.

We now discuss the parameters necessary to obtain a high precision gravitational acceleration measurement. Specifically, we need to obtain a large relative phase shift which is proportional to gravitational acceleration. From Eq.(\ref{eq:phase}), by setting a large magnetic gradient $B_g = 10^6\SI{}{T/m}$ and the oscillating period $t_0 = \SI{2}{ms} \lesssim T_1$ comparable to relaxation time of the NV, we can obtain a phase shift $\Delta \phi = \num{1.4e9}$ which is three orders of magnitude larger than phase shift in cold atom experiment $\Delta\phi = \num{3.8e6}$ \cite{Peters2001} and lead to a raise in precision by three orders of magnitude, if the errors in phase measurement is comparable. 

Why our gravimeter based on nano-mechanical matter-wave interferometry is orders of magnitude preciser than the cold atom interferometry gravimeter?
 For a diamond sphere with radius $R \sim \SI{200}{nm}$, considering the density $\SI{3000}{kg/cm^3}$ for diamond, the corresponding mass is $ m \sim \num{1e-16}\SI{}{kg}$, which is $~10^{10}$ times massive than sodium atoms, and the oscillating amplitude $\sim \SI{50}{nm}$.
  The cantilever nano-mechanical resonator is at least $\num{e16}$ times massive than sodium atoms frequently used in interferometry with much smaller oscillating amplitude $\Delta z$. Therefore, our device can be made on chip, and much smaller than atom interferometry based gravimeter. %, well below the characteristic length of the magnetic field $\sim \SI{300}{nm}$ achievable in experiments \cite{Kolkowitz2012a}. 
Rewriting phase shift Eq.(\ref{eq:phase}) using Eq.(\ref{eq:amplitude})
\begin{equation}
	\Delta \phi = \frac{16\pi m g \Delta z}{\hbar \omega_z},
\end{equation}
we can see that a large mass leads to a large phase shift within the small range of the interferometer ($\sim \Delta z$). 

Here we discuss the features of the two proposals. For the cantilever scheme, the magnetic tip can approach the NV center $<\SI{100}{nm}$ and induce a magnetic field gradient $\sim \num{1e7}\SI{}{T/m}$. The dephasing time of the NV in bulk diamond is exceptional $ \sim \SI{2}{ms}$, therefore the main decoherence effect is the mechanical damping. Recently, high $Q \sim \num{8e8}$ nanomechnical resonators using elastic strain engineering \cite{Ghadimi2017} has been realized in experiment, leading to low thermal decoherence rate $k_B T/\hbar Q$ exceeds one oscillation periods. 

For the trapped nanoparticle scheme, the coherence time of NV center can be prolonged to the limit of $T_1$ by decoherence decoupling techniques. In our proposal, no transition between $|\pm1\rangle$ is needed during the propagation, therefore it is convenient to use a continuous dynamical decoupling which prolong the quantum memory to $T_2 \sim \SI{2}{ms}$. For example, we can use the time-dependent detuning method described in \cite{Cohen2017}. Considering only the states $|\pm1\rangle$ during the propagation, the two-level system with an ambient magnetic field noise $ \delta B(t) $ is described by
\begin{equation}
	H = \frac{\omega_0}{2}\sigma_z + \delta B(t) \sigma_z.
\end{equation}
To compensate for this noise, which causes dephasing, we use a single continuous dynamical decoupling driving field
\begin{equation}\label{eq:DD}
	H_{DD} = [\Omega_1 + \delta \Omega_1(t)]\sigma_x \cos(\omega_0 t + \phi(t)),
\end{equation}
with Rabi frequency $\Omega_1 \ll \omega_0$ and Rabi frequency fluctuation $\delta \Omega_1(t)$. The time-dependent detuning
\begin{equation}
	\phi(t) = 2\Omega_2/\Omega_1 \sin \Omega_1 t
\end{equation} 
with $\Omega_2 \ll \Omega_1$. By carefully tuning the parameters, this dynamical decoupling scheme can prolong the coherence time to $T_2 \sim \SI{2}{ms}$.

As we can see from the above discussion, each of the two methods has its own advantages. The optically trapped nanoparticle has ultra-high quality factor  $Q\sim 10^{12}$,  and its trapping frequency can be tuned to optimize the phase shift. Becouse of its high Q, the proposed scheme could be performed even under room temperature. The cantilever setup does not require a  laser system to cool or trap the oscillator.  Besides, since the coupled NV is embedded in bulk solid, the dephasing time $T_2$ is significantly longer than the one in nano-diamond. Therefore, the precision of the gravimeter could be
greatly enhanced, compared with the other setup.

We have ignored various noise effects in the above estimation of phase shift. In the following section, we show that how to optimize the phase shift after considering random and systematic noises.

\section{Noise estimation}
\subsection{Fringe visibility and random noise}
The decoherence effect will reduce fringe visibility, leading to increase in amplitude noise terms (shot noise and detection noise). Here we explore two main decoherence effects: (1) motional decay of the optomechanical system; (2) dephasing of the NV center.

In the trapped particle scheme, the motional decay is associated with photon scattering from the trapping laser and heating due to random momentum kick with residual gas particles \cite{Chang2010,Romero2010,Yin2011}. The background gas collision leads to heating with a damping rate $\gamma_g/2 = (8\pi)(P/vr\rho)$ \cite{Chang2010}, where $\rho$ is the material density, $P$ and $v$ are the background gas pressure and mean speed, respectively. For a sphere of radius $R=\SI{200}{nm}$, $\omega_z = 2\pi \times \SI{0.5}{kHz}$ and a room-temperature gas with $P=\num{1e-9}\SI{}{Torr}$, the damping rate $\gamma_g/\omega_z \sim \num{4e-10}$. We define $\gamma_{sc}$ as the photon scattering induced decay rate. For the diamonds with permittivity $\epsilon = 1.5$, radius $R = \SI{200}{nm}$ and trapping wavelength $\lambda_0 \sim 10 ~\mu m$ ($\text{CO}_2$ laser), we have $\gamma_{sc}/\omega_z = (16\pi^3/15)[(\epsilon - 1)/(\epsilon + 2)]R^3/\lambda_0^3 \simeq \num{3.8e-5}$ \cite{Chang2010}, which is much larger than $\gamma_g$. Therefore the main motional decoherence comes from photon scattering with the maximum decoherence rate $\Gamma= \gamma_{sc}|2\lambda/\hbar \omega_z|^2$. With parameters in the previous section, $\lambda/\hbar\omega_z\simeq 90$, we have $\Gamma/\omega_z=0.3$ less than $1$. To further reduce the scattering noise, we could lower the maximum separation $\lambda/\hbar\omega_z$ by either reducing the magnetic gradient or increasing trap frequency. It is also possible to use other trap, e.g. ion trap \cite{Delord2017,Aranas2017}, where no photon scattering noise exists.

The other detrimental effect on the trapped nano-diamond scheme is due to the dephasing of the NV center. The noise is induced by magnetic field fluctuation of the diamond lattice and coupling to torsional motion \cite{Wan2016b}. Both effect can be suppressed by the dynamical decoupling scheme \cite{Cohen2017} which prolong the quantum memory to $T_2 \sim \SI{2}{ms}$. For $t_0 \sim T_2$, the fringe contrast is reduced to $1/e \sim 0.36$.

In the cantilever scheme, the decoherence effect can be significantly reduced. The effect of motional decay of the mechanical oscillator can be estimated by the damping rate $\gamma_{sc}/\omega_z = 1/Q \sim \num{1e-8}$ \cite{Ghadimi2017}, and the maximum decoherence in one oscillation period $\gamma_{sc}|2\lambda/\hbar \omega_z|^2 2\pi/\omega_z \sim 0.001$, which can be ignored. When the oscillator is cooled to sub-millikelvin temperature \cite{Rabl2009,Ma2016,Macquarrie2017}, the Brownian motion amplitude $\sim \SI{7}{nm}$ is much smaller than the maximum separation between the spin up and down states, thus the motional decay caused by thermal motion is negligible. The pure dephasing time of the NV center in bulk diamond could be $\sim \SI{10}{ms}$, which is much longer than the one in the trapped nano-diamond.

\begin{figure}[htbp]
	\centering
	\includegraphics[width=.95\linewidth]{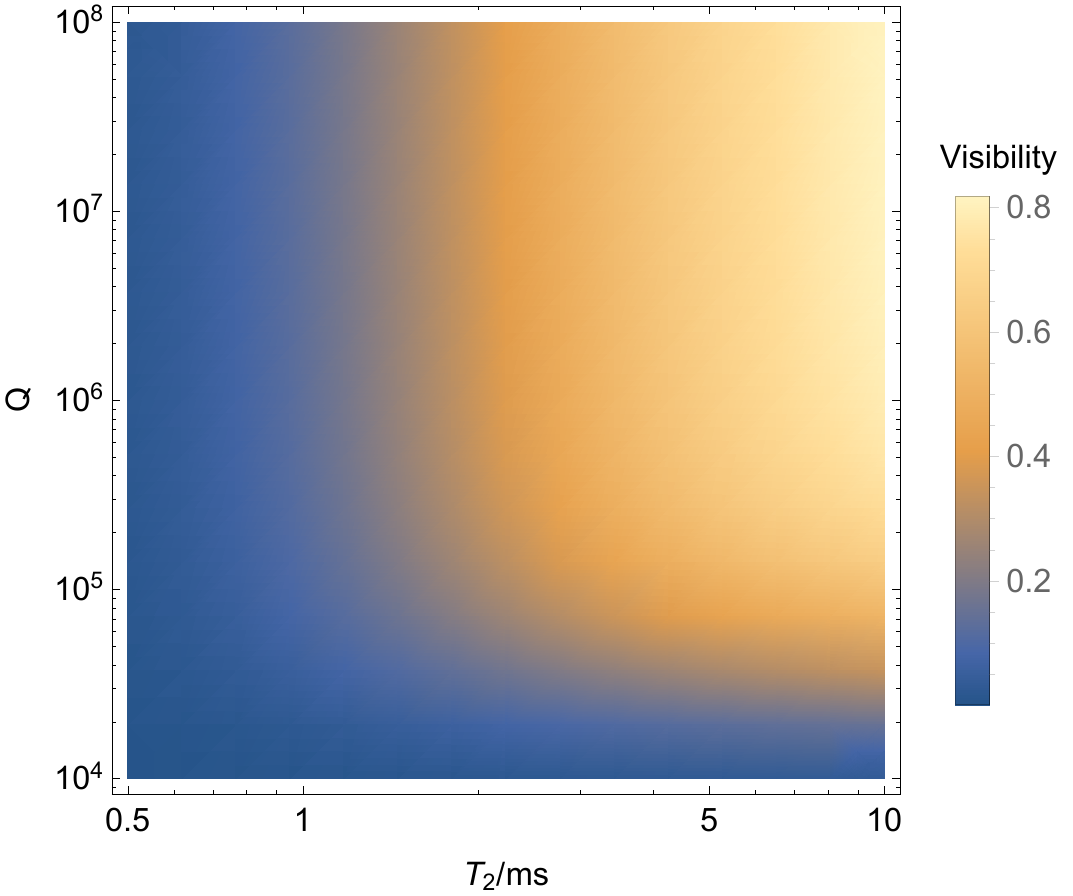}
	\caption{The visibility due to mechanical motion decoherence and dephasing under different mechanical quality factor $Q = \omega_z/\gamma_{sc}$ and the pure dephasing time $T_2$, where $\omega_z = 2\pi \times \SI{0.5}{kHz}$ and $\lambda/\hbar\omega_z\simeq 90$. The CoM motion temperature of the oscillator is cooled to $ T = \SI{0.1}{mK}$ by feedback cooling, therefore the heating effect  can be ignored.}
	\label{fig:ptvis}
\end{figure}

The visibility due to mechanical motion decoherence and dephasing under different mechanical quality factor $Q = \omega_z/\gamma_{sc}$ and the pure dephasing time $T_2$ is plotted in Fig.~\ref{fig:ptvis}, where $\omega_z = 2\pi \times \SI{0.5}{kHz}$ and $\lambda/\hbar\omega_z\simeq 90$. Here we use a simple formula $\exp(-\frac{2\pi}{Q}|2\lambda/\hbar \omega|^2)\exp(-t_0/T_2)$ to estimate the visibility. To retain a large visibility, the quality factor Q should be larger than $\num{1e5}$ and the dephasing time $T_2$ should be longer than $\SI{2}{ms}$.

The Ramsey fringe visibility is directly related to the signal-to-noise ratio, and will appear in the amplitude noise terms. To reduce the shot noise limit to the measurement precision $\Delta g/g$ below $\num{1e-10}$, we need at least $\num{1e5}$ data points for each measurement. For the atom interferometry, there are $N \sim \num{1e6}$ atoms in the atom fountain simultaneously contribute to the signal. As for our scheme, we should fabricate $M$ mechanical resonators with the same frequency on-chip, which can perform the measurement at the same time. With the measurement repeating frequency ~$1$ kHz and $M=100$, we can achieve the precision goal within $\SI{2}{s}$. We note that  modern technology makes it possible to fabricate such solid-based gravimeter on chip.

\begin{figure}[htbp]
	\centering
	\includegraphics[width=\linewidth]{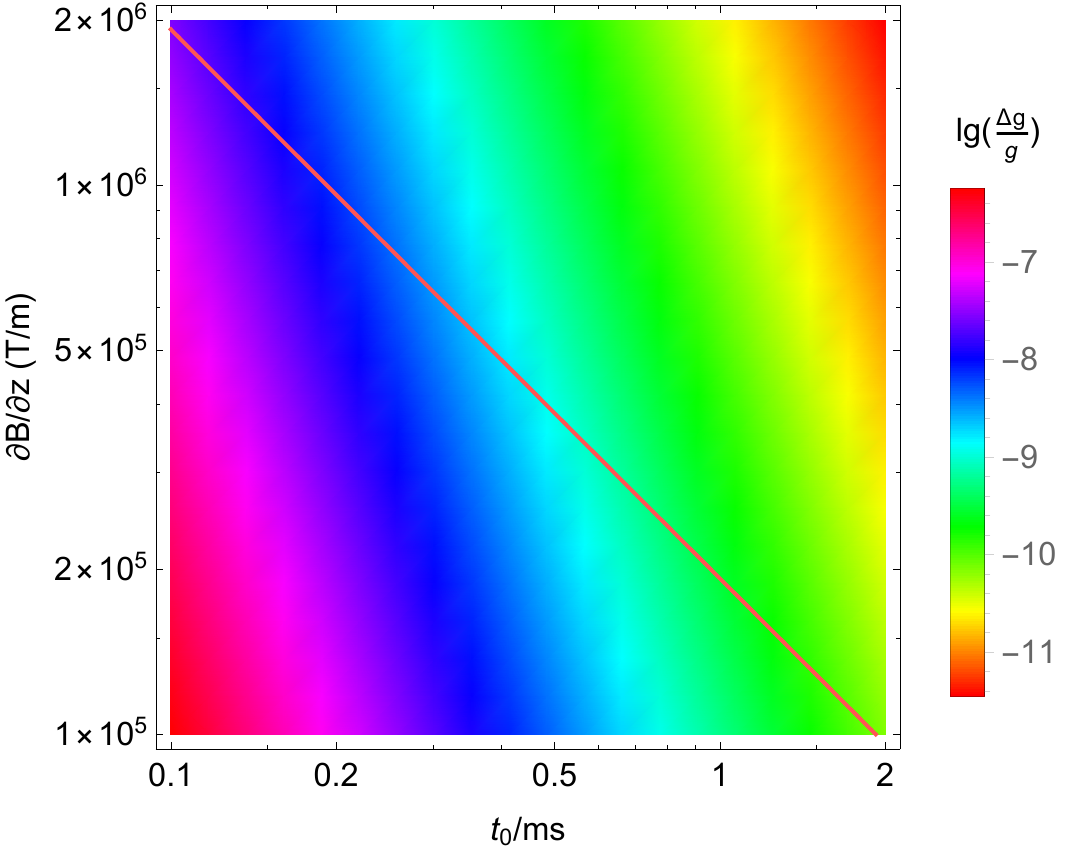}
	\caption{Precision of the gravity acceleration measurement under different magnetic field gradient $B_g$ and oscillation period $t_0$. We chose $t \le \SI{2}{ms}$ and assume the visibility are high enough by choosing the parameters to be in the upper right region in Fig.\ref{fig:ptvis}. We take the accuracy of the phase shift to be $\SI{10}{mrad}$ in the estimation. In addition, we require the magnetic field fluctuation to be $\Delta < \SI{10}{MHz}$, which corresponds to the region below the red line. Within the region, up to \num{1e-10} relative precision can be achieved at the lower right corner.}
	\label{fig:prec}
\end{figure}

We assume that the phase shift can be measured with precision $\SI{10}{mrad}$, comparable to atom interferometry. Based on Eq.\eqref{eq:phase}, we can plot the precision of our gravimeter under different experiment parameters, as shown in Fig.~\ref{fig:prec}. In order to retain a high fringe visibility, the choice of parameters is limited to the region below the red line. This is because under  the external magnetic gradient, the thermal motion of the oscillator leads to a magnetic field fluctuation for the NV center and the dephasing of the NV center electron spin. The fluctuation in magnetic field can be estimated by the root-mean-square of the mechanical motion times the magnetic gradient, which reads
\begin{equation}
	\Delta = B_g \sqrt{\frac{k_B T}{m \omega_z^2}}
\end{equation}
If the condition $ \Delta \ll \Omega_1$ fulfills, the extra dephasing can be suppressed by dynamical decoupling drive with Eq.\ref{eq:DD}. This condition limit our  choice of the parameters. We assume that the center of mass motion temperature of the trapped nano-diamond is cooled to $ T = \SI{0.1}{mK}$ by feedback cooling, and the mass of the nanoparticle is $m = \num{1e-16}\SI{}{kg}$. For typical Rabi frequency $ \Omega_1 \lesssim \SI{100}{MHz}$, we require the magnetic field fluctuation to be $\Delta < \SI{10}{MHz}$, which corresponds to the region below the red line in Fig.\ref{fig:prec}. If the CoM temperature could be cooled to $T < 1~\mu$K, e.g. by cold atom \cite{PhysRevA.91.013416}, the measurement precision would be further improved by one to two orders of magnitudes.

\subsection{Systematic noise}
Thanks to the short matter-wave wave length of both the nanoparticle and cantilever, our scheme has the advantage of on-chip compared with the atomic gravimeter. Therefore, the laser system, vibration isolator
and other auxiliary devices shoulde be relatively easy to reach a high
precision.  We assume that in our setup the systematic error is of the same order of magnitude as the one in atom interferometry method. 
%Systematic errors in the instrument include uncertainty in the rf phase shift, the Coriolis effect, the wavelength of the caesium D1 transition to which the frequencies of the Raman lasers are referenced, the laser lock offset, gravitational field gradients, ac Stark shift and precise pulse timings \cite{Godun2001}. Nevertheless, 
The resulting interferometer is accurate enough to allow phase shifts $\SI{10}{mrad}$ to be detected, leading to a value of $g$ accurate to one part in \num{1e10} in our method. 

The second order magnetic field gradient induces extra systematic error. The additional term in Lagrangian is  $ \pm\frac{1}{2}g_{NV}\mu_B\frac{\partial^2 B}{\partial z^2}z^2$, which is different for spin up and down. We can treat it as a shift in trapping frequency $\Delta \omega_{\pm}/\omega = \pm \frac{1}{8 m}g_{NV}\mu_B\frac{\partial^2 B}{\partial z^2}$. The frequency error contribute to the measured gravitational acceleration $\frac{\Delta g}{g} \sim \frac{\Delta \omega}{\omega}$. By this relation, we get the maximum second derivative of the magnetic field $\frac{\partial^2 B}{\partial z^2} \lesssim \num{1.7e5} \SI{}{T/m^2}$ for $\frac{\Delta g}{g} = \num{1e-10}$. In the case of cantilever oscillator, we can embed another magnetic tip under the NV in the bulk diamond to reduce $\frac{\partial^2 B}{\partial z^2}$. For trapped nanoparticle scheme, we can eliminate this effect by simultaneously rotating the trapping direction $\ang{180}$ and apply a microwave $\pi$-pulse to flip the spin states after one evolution period $t_0$ and measure the phase shift at time $2t_0$. Because  the trapping axis is rotated, the phase shift of the two evolution period accumulates, while the second order magnetic field induced phases will be cancelled out.

Other systematic noise includes magnetic field drift, perturbative terms related to $\omega_x,\omega_y$, anharmonic effects of the trapping potential, random orientation and Doppler effect. The magnetic field drift in a timescale of hours is much larger than the system evolution time of $\SI{1}{ms}$. So in principle, the magnetic gradient can be determined with relative accuracy $e^{-t/t_{drift}} \sim \num{1e8}$ or higher if we consider its time dependence. Perturbative analysis shows that for $\omega_x = \omega_y = 10 \omega_z$, the fidelity of the evolution stays above $99\%$ even when the initial state is thermal with an average thermal occupation number is up to 600 \cite{Wan2016}. The anharmonic effects of the trapping potential will be avoided by feedback cooling of our oscillator to sub-millikelvin temperatures. The random orientation effect can be corrected by methods shown in \cite{Cohen2017}. The Doppler effect may appear in the spin preparation when the nanoparticle is oscillating in the trap. The corresponding frequency error could be estimated by $\delta f = f_0 v/c = f_0 \Delta z \omega_z/c$, where $\Delta z$ is the oscillation amplitude, $f_0 = \SI{2.88}{GHz}$ is the microwave frequency in use, and $c$ is the speed of light. With $\delta z \approx \SI{100}{nm}$ and $\omega_z \approx 2\pi \times \SI{1}{kHz}$, we eventually have $\delta f \approx \num{6e-3}\SI{}{Hz}$, which is much smaller than the typical linewidth of an NV center of $\sim\SI{10}{MHz}$. For a thermal state of the CoM, with a temperature cooled to about $\SI{1}{mK}$ in the trap, the root-mean-square velocity is about $v_1 = \sqrt{2kT/m} \sim \SI{0.002}{m/s}$. Therefore the Doppler shift would not be a concern in our scheme.

\section{Conclusion}

In conclusion, we have proposed a solid-base on-chip gravimeter which makes use of the matter-waver interference of a mechanical resonator to significantly increase the precision.
 We have proposed two equivalent schemes to couple an NV center to a mechanincal oscillator in gravitational field. In order to measure the gravitational acceleration, we have proposed the method to achieve Ramsey interferometry in this system, where the inference pattern is depend on the gravitional induced phase shift. 
 Under the experimental feasible parameters, we found that the phase shift is three order of magnitudes greater than the atomic interferometry method.
We then analyze the noise effects, including motional decay of the oscillator and the dephasing of the NV center, on the precision of the gravimeter. It is found that 
the relative precision $10^{-10}$ is possible under the current experimental conditions.
Finally, we have analyzed the effect of the second derivative of the magnetic field and provided methods to compensate it.

\begin{acknowledgments}
Z.Q.Y. is supported by National Natural Science Foun-
dation of China NO. 61771278, 61435007, and the Joint
Foundation of Ministry of Education of China (6141A02011604).
We thank the helpful discussions with Huizhu Hu and Tongcang Li.

\end{acknowledgments}

%\bibliography{gravMetry}

\end{document}